# Hybrid ML-RL Approach for Smart Grid Stability Prediction and Optimized Control Strategy

Kazi Sifatul Islam
Ingram School of Engineering
Texas State University
San Marcos, USA
kazi_sifat@txstate.edu

Anandi Dutta
Ingram School of Engineering
Texas State University
San Marcos, USA
myb43@txstate.edu

Shivani Mruthyunjaya
Ingram School of Engineering
Texas State University
San Marcos, USA
aon9@txstate.edu

*Abstract*— Electrical grids are now much more complex due to the rapid integration of distributed generation and alternative energy sources, which makes forecasting grid stability with optimized control a crucial task for operators. Traditional statistical, physics-based, and ML models can learn the pattern of the grid features, but have limitations in optimal strategy control with instability prediction. This work proposes a hybrid ML-RL framework that leverages ML for rapid stability prediction and RL for dynamic control and optimization. The first stage of this study created a baseline that explored the potential of various ML models for stability prediction. Out of them, the stacking classifiers of several fundamental models show a significant performance in classifying the instability, leading to the second stage, where reinforcement learning algorithms (PPO, A2C, and DQN) optimize power control actions. Experimental results demonstrate that the hybrid ML-RL model effectively stabilizes the grid, achieves rapid convergence, and significantly reduces training time. The integration of ML-based stability classification with RL-based dynamic control enhances decision-making efficiency while lowering computational complexity, making it well-suited for real-time smart grid applications.

I. INTRODUCTION

Global energy demand is evolving rapidly worldwide, and to mitigate the demand, most countries use and produce many different types of energy sources, such as nuclear energy, fossil fuels (petroleum, natural gas, and coal), and renewable energies (wind, solar, geothermal, biomass, hydropower) which are primary energy sources [1]. However, electricity is a secondary energy source generated from primary sources [2]. To convey electricity to the energy-use sectors and maintain the demand and supply balance, a well-established grid system is required, which is nothing but a complex network that contains three major parts, including generation, transmission, and distribution. Integrating renewable energy sources into the grid affects grid stability and is mainly for frequency and voltage control, variability and intermittency, demand response, flexibility, and energy storage solutions. Considering different features related to the grid operation, forecasting grid stability is one of the significant challenges in the current energy industry. Many methods and devices meet these conditions and achieve high efficiency and stability. In a smart grid, consumer demand information is collected and centrally evaluated against current supply conditions, and the resulting proposed price information is sent back to customers to decide about usage. Addressing and simplifications in the grid model, a mathematical approach including different equation-based models can be considered in several ways to predict stability; however, the execution of this model relies on significant simplifications [2]. To overcome the inherent DSGC (Decentralized Smart Grid Stability) model simplifications, many alternative approaches, such as different machine learning models like Support Vector Machine (SVM), Decision Tree, XGBoost, etc., have been proposed and implemented. Complex nonlinear relationships, high dimensionality, and feature extraction on time series sequential data handling limitations are the pivotal drawbacks of these methods. Neural network-based deep learning approaches provide more adaptability and performance benefits for predicting grid stability and overcoming challenges. Electricity demand and other features tend to peak over a specific time under particular weather conditions; neural network models might learn that phenomenon, which can be encoded in the weights of the model's neurons. Reinforcement Learning (RL), on the other hand, offers a promising approach to real-time control by continuously learning optimal policies for stabilizing the grid. RL agents interact with the environment, optimizing power control based on rewards associated with system stability. Despite its potential, RL alone suffers from slow convergence, high computational demands, and instability in training due to extensive exploration requirements.

This paper presents a hybrid ML-RL approach that combines the strengths of both machine learning (ML) and reinforcement learning (RL) paradigms. The ML model is a rapid classifier that predicts grid stability conditions, while the RL model fine-tunes power control actions when instability is detected. Integrating ML with RL enhances learning efficiency by filtering stable states, reducing the RL model's search space, and accelerating decision-making.

## II. LITERATURE REVIEW

Different electric grid stability systems have been proposed and implemented for optimal grid control and instability prediction. Previous work on this problem can be categorized mainly into traditional statistical and physics-based techniques, Machine Learning and Deep Learning approaches, and Reinforcement Learning techniques.

### A. Traditional Approaches

Several traditional approaches have been used in predicting grid stability, including rule-based and deterministic models such as state-space models, vector computation, power flow analysis, etc. K.I et al. proposed a strategy consisting of using a genetic eigenvalue technique to generate eigenvalues, damping ratios, and participation factors for proper placement of PSS (Power System Stabilizers) to mitigate the effect of transmission line and power plant outage contingencies [3]. In another article by Xie et al., along with the eigenvalue analysis, the system's stability was judged according to the minor signal stability analysis criterion. The system stability under the interference of a stochastic small signal is analyzed, and an analytical method to judge the stability based on the mean stability criterion and the mean square stability criterion is presented [4]. Bombois et al. proposed a system identification-based approach to monitor and redesign Power System Stabilizers (PSS) without requiring dynamic models, leveraging real-time measurements and "probing" technologies to improve damping control and system stability [5]. The small signal stability analysis and eigenvalue analysis method, which has certain drawbacks in that it fails to capture the intricacies and non-linearities of real-world stochastic events adequately, is the exclusive focus of these works.

### B. Machine Learning and Deep Learning Approaches

Leveraging the ability to model complex, nonlinear relationships and handle multidimensional data, many supervised, unsupervised, and even semi-supervised machine learning (ML) and deep learning (DL) models play a significant role as alternatives to traditional approaches. For example, Fateh Karim et al. contribute to the field of intelligent grid stability prediction by introducing an optimized Long short-term memory (LSTM) model enhanced by the Novel Guide-Waterwheel Plant Algorithm (Guide- WWPA), demonstrating superior performance in capturing temporal dependencies and nonlinear relationships in data [6]. Implementing the Guide- WWPA optimization, generalizability, and computational costs affect the model's purpose. Gunjal et al. introduced a Light Gradient Boosting Machine (LGBM) model that achieves a high accuracy of 96 percent in predicting grid stability, outperforming other classifiers like SVM and Decision Tree [7]. In another study, Marevac et al. explore the application of neural networks and deep learning for grid stability predictions. They aim to use predictive models to assess voltage stability, detect potential instability events, and optimize control strategies for effective stability control measures [8]. The dataset used for evaluation is sourced from the IEEE 14-bus system, which serves as a standard benchmark in power system studies, allowing for effective validation of the proposed models. However, the reliance on a single dataset may limit the generalization of the findings. A proactive semi-supervised method was also proposed in this IEEE 14-bus system, where Abegaz et al. suggested a proactive semi-supervised machine learning method using a graph model for stability estimation in smart grids, achieving a high correlation between predicted and actual terminal voltages in five test cases, including the IEEE 14-118 bus systems [9]. A Convolutional Neural Network (CNN) design for grid stability analysis was also proposed by Mewada et al. in their study. This CNN effectively captures spatial and temporal dependencies in grid data, enabling accurate prediction and classification of stability-related events with high accuracy, precision, recall, and F1 score [10].

### C. Reinforcement Learning Approaches

All these traditional ML and DL approaches have been proposed for predicting instability, but have not fully explored the strategies for transforming it from "unstable" to "stable." Many reinforcement techniques have been proposed to optimize the grid's stability, ranging from RL agents' efficacy in response to adverse grid events to primarily relying on the load balancing of the grid. A study employs the proximal policy optimization (PPO) algorithm and the graph neural networks (GNNs). They used the Grid2PO platform for agent training, and the agent's performance is expressed concisely through its reward function [11]. Another study proposes an RL approach using the Markov Decision Process (MDP) framework for dynamic load scheduling in electric grids. They investigate applying basic algorithms, such as Q-Learning, to more advanced Deep Q-Networks (DQN) and Actor-Critic methods [12]. By taking into account the grid safety, unit operation, and power balance, a suitable regional power grid simulation platform and deep reinforcement learning (DRL) approaches are also proposed by a study where action and state spaces are consistent with the operational adjustments of equipment in the power grid [13]. Apart from these, the multi-agent reinforcement learning technique was also utilized in building energy management systems for grid awareness; they consider voltage regulation on the IEEE-33 bus network of controllable loads, energy storage, and small inverters of the building. They found that the RL agent nominally reduced the undervoltage and overvoltage scenarios by 34% [14].

All the approaches described in this section can indicate a prospective gap in addressing the non-linearity and high-dimensionality of features of predicting grid stability along with the stability control strategy. We are focusing on bridging the gap between stability prediction and actions for stability control using hybrid RL-ML methods and determining the potential of reinforcement algorithms.

## III. SYSTEM MODEL AND PROBLEM FORMULATION

The system under consideration is a simulated 4-node star topology electrical grid, consisting of one central generation node and three consumer nodes. Each node is characterized by dynamic parameters such as power consumption or production, response time, and elasticity to price changes. This configuration reflects a simplified yet representative model of real-world smart grid environments where node-level behaviors significantly impact system-wide stability.

In such grids, stability is frequently challenged by fluctuating demand, the integration of intermittent energy sources, delayed reaction times, and highly elastic user behaviors. These factors introduce non-linear dynamics and uncertainty into grid operations, making it difficult to maintain a stable state through traditional control methods. As demand patterns shift rapidly and nodes respond asynchronously, the system can become unstable, leading to operational inefficiencies or failures. The central research problem addressed in this study is twofold: first, to accurately predict whether the grid is in a stable or unstable state using node-level data; and second, to identify a control strategy that can dynamically respond to instability and guide the system back to a stable condition.

To address this, we propose a hybrid approach that combines machine learning for real-time stability prediction with reinforcement learning for adaptive control. This framework enables efficient decision-making by detecting instability early and responding with optimized control actions to enhance grid resilience and operational efficiency.

### A. Features

The original dataset contains 10,000 observations. As the reference grid is symmetric, the dataset was augmented 3! (3 factorial) times, or 6 times, representing a permutation of the three consumers occupying three consumer nodes. The augmented version has then 60,000 observations and contains 12 primary predictive features and two dependent variables. The features of the datasets are the reaction time of each network participant, which has to be the actual value within the range 0.5 to 10, power produced (positive) or consumed (negative) by each network node, an absolute value within the range -2.0 to -0.5 for consumers, price elasticity coefficient for each network participant, an absolute value within the range 0.05 to 1.00.

### B. Exploratory Data Analysis

Before the advent of any preprocessing and model evaluation, we run through an EDA process to check the inner pattern of the dataset. Fig. 1 demonstrates the correlation matrix of the features to find the relationship between the features and the inherent importance of every feature to each other, as well as with the target feature.

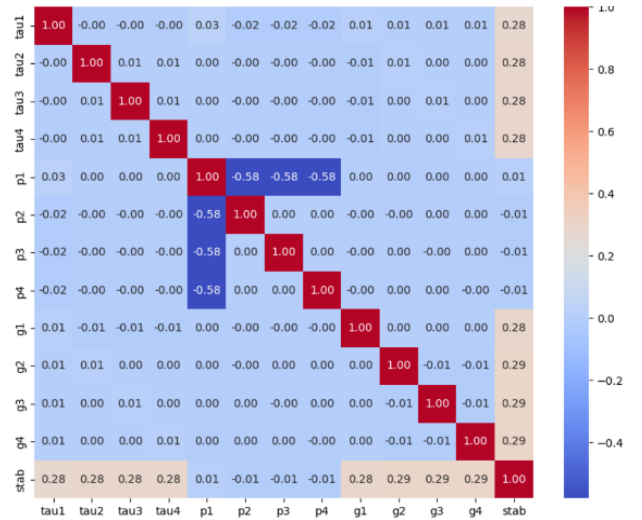

Fig. 1. Correlation matrix of the features.

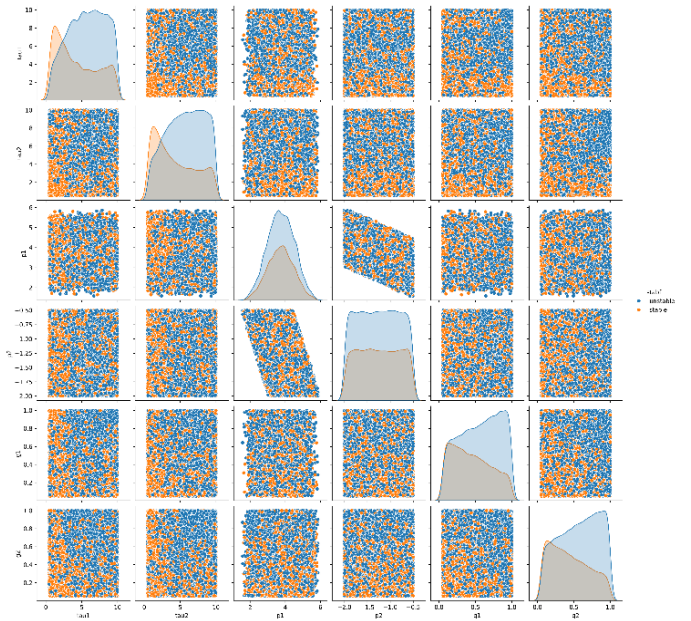

Fig. 2. Scatter plot of the feature relation with the target variable.

Fig. 2 depicts the scatter plots of each feature distribution, indicating that the dataset presents a multidimensional and non-linear pattern.

## IV. METHODOLOGY

This proposed hybrid ML-RL system consists of two main stages. In the first stage, ML approaches create the baseline and predict the "stable" and "unstable" conditions, and in the second stage, RL agents control the grid from "unstable" to "stable." Fig. 3 demonstrates the workflow of the proposed method.

### A. Part-1: ML-Based Stability Prediction

The first stage categorizes the grid state as stable or unstable using a stacking classifier. The classifier is trained on historical data from grid operations using characteristics like power flow, reaction times, and gamma function values.

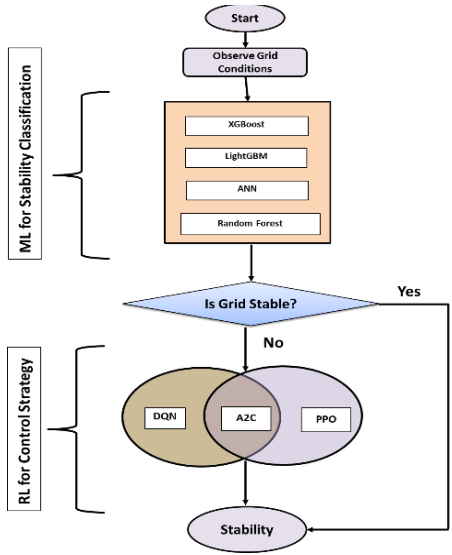

Fig. 3. Workflow chart of the Hybrid ML-RL approach.

The ML model is trained using an ensemble method where multiple base learners—Random Forest, XGBoost, ANN, and LightGBM—contribute to predictions, and a meta-classifier refines the final decision. This enables the model to achieve high predictive accuracy while reducing false positives and negatives. Fig. 4 shows the workflow of the method, and its primary benefit is its quick inference time, which instantly allows stability evaluation without requiring intricate physics-based simulations.

1) Random Forest: A random forest classifier is the first and foremost option for successfully classifying predictions, as it combines multiple decision trees to make robust predictions. It is an ensemble approach that mitigates the impact of individual trees by an aggregating or majority voting approach. Though the data has no significant outliers, Random Forest is still a perfect fit for this due to its bootstrap aggregating method during training. Training each tree on a different subset of the data makes the model less sensitive to individual data points and improves its ability to generalize to unseen data. One of the other significant advantages of Random Forest in this study is its feature importance insights.

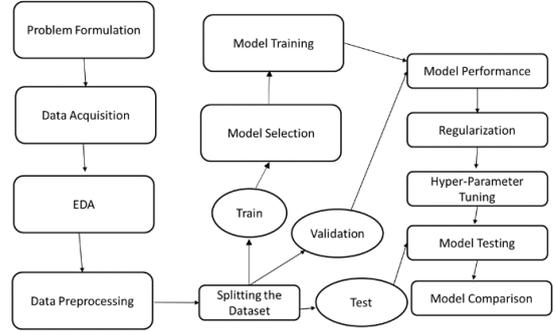

Fig. 4. Block diagram of the workflow of stage 1 (ML for prediction)

$$\hat{y} = mode\{T_1(x), T_2(x), \dots, T_N(x)\} \quad (1)$$

$$P(y = c|x) = \frac{1}{N} \sum_{i=1}^{N} P_i \ (y = c|x) \quad (2)$$

$$\hat{y} = \arg\max P(y = c|x) \quad (3)$$

Equation (1) represents the majority voting technique among N different trees, denoted as $T_1(x)$, $T_2(x)$, to $T_N(x)$, where the mode function takes the most frequently occurring prediction among these trees to find the final predicted loss $\hat{y}$. Equation (2) represents a soft voting ensemble technique by averaging the probabilities of all $N$ trees. Equation (3) indicates a rule-based probabilistic model to predict the final output where $\arg\max$ is a function used for identifying the class c that maximizes $P(y = c|x)$.

2) *XGBoost*: To make accurate predictions and learn the complex relationship of the features, XGBoost works by building trees sequentially. Each new tree corrects the errors of the previous one, and the process continues recurrently, which allows us to address the limitation of handling non-linearity in the data. The robust dataset also helps enhance speed and efficiency through parallel processing and various optimization techniques to reduce training time. It also helps to address the issue as our dataset has an imbalanced target feature distribution, as shown in Fig. 3.

3) *LightGBM*: LightGBM, developed by Microsoft, is an open-source model for this classification problem. After defining all the parameters, including the learning rate, number of iterations, etc., we tried to fit the model in our dataset to find the best accuracy for predicting grid stability. One of the most significant reasons for selecting the model is its efficient data structure for storing the trees, which results in lower memory usage. This benefits training models on large datasets and working with limited computational resources.

4) *ANN:* A fully connected Artificial Neural Network approach taken for sequentially checking the different patterns of the dataset by densely connecting the hidden layer between them. The architecture has an input layer with 13 nodes followed by two hidden layers, 64 and 32 nodes, respectively. Finally, it has one output node, making 3009 trainable parameters in the architecture. Two dropout layers with a value of 50% each were implemented after two hidden layers were implemented for regularization to prevent overfitting. The dropout layers randomly drop out the fractions of the neurons during training, making the model less dependent on specific neurons. The "Relu" activation function has been used for the hidden layers, and the "sigmoid" activation function has been used for the output layers, as the output is a binary classifier.

5) *Stacking*: Combination of the prediction of the previously described base model and 5-fold cross-validation, stacking creates a powerful ensemble learning technique and a higher-level model called meta learner; unlike bagging or boosting, stacking focuses on combining diverse models rather than iterations of a single algorithm, making it a powerful approach for complex tasks like in our dataset. As the base models are diverse and are significant in predicting accuracy, this technique can improve the overall performance and mitigate the impact of individual errors. After bagging all the base learners, the Logistic Regression technique has been used as the final estimator of this meta-learner in this stacking classifier.

B. *Part-2: RL-Based Control Optimization*

Once instability is detected, an RL agent is activated to take corrective actions in real-time.

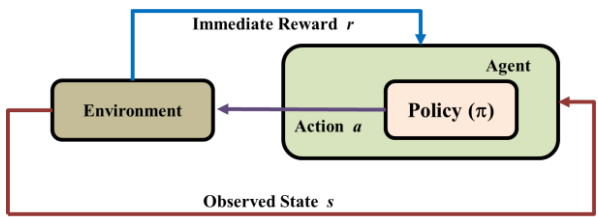

Fig. 5. Agent and Environment interaction for RL approach

The RL agent is trained in a simulation environment through a reward-based feedback system and trial-and-error learning. Positive rewards are given for any action that brings the grid closer to stability, while negative rewards are given for actions that cause instability. Fig. 5 shows the agent-environment interaction flow. When ML and RL are combined, the system is optimized to predict stability with accuracy and take prompt corrective action when needed.

1) *RL Environment Setup*: With the help of the RL environment's simulation of power grid dynamics, the agent interacts with the environment and learns from grid fluctuations. The environment design is organized as follows:

   *a) State Space:* During the reinforcement learning process, the state space is the set of all possible conditions or configurations of the grid that the agent can observe and interact with.

   $$S_t = \{\tau_1, \tau_2, \tau_3, \tau_4, p_1, p_2, p_3, p_4, g_1, g_2, g_3, g_4, stab\}$$

   The system's state space at time step t is defined as a 13-dimensional vector that encapsulates key grid parameters and represents the smart grid's current condition. Here, reaction times ($\tau_1, \tau_2, \tau_3, \tau_4$) represent the time intervals within each node response to changes in grid power flow, power consumption ($p_1, p_2, p_3, p_4$) indicates the amount of power consumed by every node, gamma function values ($g_1, g_2, g_3, g_4$) represents the dynamic response of the grid under varying conditions, and the grid stability indicator (stab) indicates whether the grid is currently stable or not.

   *b) Action Space:* The action state $A_t$ defines a set of all possible actions an RL agent needs to take to adjust the power for grid stability.

   $$A_t = \{a_1 = 0, a_2 = 1, a_3 = 3\}$$

   Agents act according to three distinct actions based on power fluctuation. The actions are labeled $a_1 = 0$, $a_2 = 1$, and $a_3 = 2$ for Decrease, Maintain, and Increase Power. After every state, the agent decides its action according to the reward function. These actions dynamically reach stability from an unstable grid condition classified by the base ML models.

   *c) Reward Function:* The agents have been selected and iteratively tested on the unseen data to maximize the learning from the training data. An optimal and well-defined reward function has been chosen for them.

   $$R(a_t, S_t) = \begin{cases} +20, & if\ stabnew > stabold \\ -20, & if\ stabnew < stabold \\ 0, & if\ stabnew \approx stabold \end{cases}$$

The reward function encourages the agents to learn the environment according to the reward and penalty values. Balancing rewards and penalties helps an agent learn a stabilization policy. The training process follows the Bellman Equation, which is shown in equation (4), where the agent updates its Q-values based on the reward received:

$$Q(s,a) = Q(s,a) + \alpha[R(s,a) + \gamma maxQ(s',a') - Q(s',a')] \quad (4)$$

Here, α is the learning rate and γ is the discount factor that determines the future rewards' importance.

1) *Algorithms Used*: The fundamental classification of the RL algorithms is Value-based, Model-based, and Hybrid of both of them. We considered an algorithm from all three categories and used it to learn the agent in the environment on our training data. The algorithms' functionalities are:

   *a) Proximal Policy Optimization (PPO):* A policy gradient-based RL algorithm that refines decision-making through clipped updates, balancing exploration and exploitation. The PPO objective function is given by the equation (5):

$$J(\theta) = \mathbb{E}[\min(r_t(\theta)A_t, clip(r_t(\theta), 1 - \epsilon, 1 + \epsilon)A_t)] \quad (5)$$

   Here, $r_t(\theta)$ is the probability ratio, and $A_t$ is the advantage function.

   *b) Advantage Actor-Critic (A2C):* A synchronous actor-critic algorithm that reduces training variance and improves stability in control policies. The policy update is computed as per the equation (6):

$$\nabla_\theta J(\theta) = \mathbb{E}[\nabla_\theta \log \pi_\theta(a_t|s_t)A_t] \quad (6)$$

   Were, $\nabla_\theta J(\theta)$ is the gradient of the objective function, representing the direction in which we update the policy parameters to maximize expected returns.

   *c) Deep Q-Network (DQN):* DQN is a value-based Q-learning algorithm that follows the Bellman Equation indicated in Equation (1). The agent receives a state representation of power, reaction times, and gamma function values. These inputs determine the best action that can be performed for stability. It follows the experience replay phenomena for storing and sampling the learning efficiency. Q-value divergence prevention and stability enhancement done by the DQN target network. Table I. lists the optimized hyperparameters used in all value and model-based approaches for the RL.

TABLE I: HYPERPARAMETERS OF ALL RL ALGORITHMS.

| Parameter | PPO | A2C | DQN |
|---|---|---|---|
| Learning Rate | 0.0003 | 0.0003 | 0.0003 |
| Discount Factor (γ) | 0.99 | 0.99 | 0.99 |
| Batch Size | 64 | N/A | 64 |
| Policy Type | MlpPolicy | MlpPolicy | MlpPolicy |
| Exploration Strategy | Adaptive | Adaptive | Q-Learning |

V. RESULTS AND COMPARISONS

This section evaluates the performance of our two-part hybrid ML-RL framework. In the first part, multiple machine learning models are tested for their ability to classify grid states as stable or unstable. The best-performing model is selected based on standard classification metrics and compared with existing studies to highlight its effectiveness. In the second part, reinforcement learning models are applied only when the ML model detects instability. These RL agents learn optimal control actions to restore grid stability, and their performance is analyzed in terms of success rate, convergence speed, and training efficiency. The final hybrid model combines the top ML and RL components to deliver accurate predictions and optimal stabilization.

*A. Comparative Analysis of ML-based Prediction Stage*

The performance metrics of five different models, including Random Forest, XGBoost, LightGBM, ANN, and the Stacking model, are presented in Fig. 6. Key evaluation metrics considered in this study are precision, recall, and F1-score for both the "stable" and "unstable" classes. As shown in the bar chart, all models exhibit high and consistent performance, with most scores ranging between 0.93 and 0.99 across both classes. The Stacking Ensemble model stands out as the top performer, achieving 0.97 or higher in every metric, including an impressive F1-score of 0.98 for both classes. XGBoost and LightGBM also perform strongly, with scores hovering around 0.96–0.98, while ANN and Random Forest follow closely, each maintaining precision and recall values above 0.93 in all metrics. These results confirm that all models are capable of reliable grid stability prediction, with the Stacking model demonstrating the most balanced and robust classification capability.

This outstanding performance highlights the Stacking model's effectiveness in reducing both false positives and false negatives. Based on the results shown in Fig. 6, we selected the Stacking classifier for the first part of our hybrid ML-RL model to detect instability with high confidence before triggering the RL control phase. We further compare our prediction accuracy with other recent studies on the same dataset in Table II.

The models used in those works include various well-known algorithms such as XGBoost, Ensemble Bagging, and different ensemble learning techniques. Our study surpasses these previous efforts by achieving the highest accuracy on the augmented version of the UCI smart grid dataset [15].

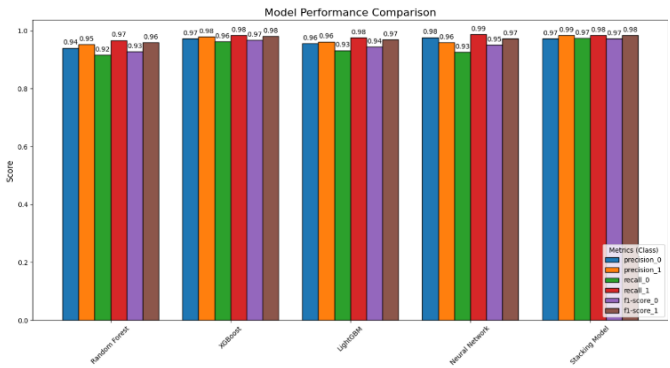

Fig. 6. Performance comparison of RF, XGBoost, LightGBM, ANN, and Stacking classifier for the stage (Stability prediction).

Specifically, our XGBoost model achieved an overall accuracy of 97.21%, outperforming the 94.70% accuracy reported in a 2022 study that used the same dataset without augmentation. LightGBM followed closely with an accuracy of 97.98%, and the Random Forest model achieved the highest accuracy of 98.21% among all individual models. The ANN model reached an accuracy of 96.91%, still demonstrating strong classification capability. Most notably, though our stacking classifier has a tiny lower accuracy than the Random Forest and LightGBM, it reached an F1-score, precision, and recall that are superior to others for both stability and instability predictions, making it the most balanced and robust among all tested models.

TABLE II: COMPARISON WITH OTHER STUDIES IN SIMILAR DATASETS.

| Study | Dataset Instances | Models | Accuracy |
|---|---|---|---|
| 2022[16] | 10000 | XGBoost | 94.70% |
| 2022[17] | 10000 | Ensemble Bagging | 90.16% |
| 2023[18] | 10000 | Cross-sensitive stacked ensemble classifier | 96.25% |
| 2023[19] | 10000 | MLP-Extreme Learning Machine | 95.86% |
| 2024[20] | 60000(Augmented) | Ensemble Stacking Classifier | 97.77% |
| | | Ensemble soft voting classifier | 95.23% |
| | | Ensemble hard voting classifier | 95.26% |
| This study | 60000(Augmented) | Random Forest | 98.21% |
| | | XGBoost | 97.21% |
| | | LightGBM | 97.98% |
| | | ANN | 96.91% |
| | | Stacking Classifier | 97.88% |

Table II also highlights the impact of dataset augmentation; expanding the original dataset from 10,000 to 60,000 instances significantly improved model performance. Even with ensemble approaches, prior studies using the base dataset could not match the accuracy and class-level precision our model achieved on the augmented version. These results and comparisons depict that our study set up a new benchmark compared to previous research by combining augmented data with advanced preprocessing and ensemble methods.

### B. Performance Analysis of RL-based Control Stage

After classifying the unstable condition through the best-performing model, the stacking classifier, we consider different aspects, including the success rate, training time, convergence speed, and reward function progression over training, to analyze the performance of RL methods (PPO, A2C, and DQN). The success rate in this stage reflects how effectively each RL model restores grid stability after instability is predicted by the stacking classifier in the ML stage. A bar chart in Fig. 7 illustrates the success rates of the reinforcement learning models used for restoring grid stability after instability is detected by the stacking classifier. Among the three RL agents, PPO achieved a 98% success rate, meaning it was able to successfully stabilize the grid in 98 out of 100 test episodes. The small gap from perfection may stem from PPO's clipped policy updates, which balance exploration and exploitation but can occasionally hinder optimal decision-making in highly sensitive or borderline cases. In contrast, both A2C and DQN achieved 100% success rates, consistently stabilizing the grid in all tested conditions. These results highlight the reliability of value-based (DQN) and actor-critic (A2C) approaches in executing precise control actions under dynamic grid conditions. However, while success rate is a critical metric, it does not reflect the speed of convergence or the computational efficiency of each model. Therefore, it is also necessary to evaluate training time and convergence speed to determine which hybrid pipeline offers the best performance for real-time smart grid applications.

Training time depends on several factors, such as the number of episodes the model needs, batch size, learning rates, and environmental complexity. In this aspect, Fig. 8 shows a bar chart comparing the relative training times of the reinforcement learning agents used for stabilizing the grid following instability classification by the ML model. Among the three, PPO required the longest training time relative to the baseline, due to its complex policy optimization steps and extensive exploration strategy. A2C exhibited moderate training time, benefiting from synchronous updates but still requiring more iterations to stabilize compared to simpler value-based methods. DQN was the most efficient, completing training in the shortest time, aided by its use of experience replay and a fixed target network, which streamlines the learning process. These findings suggest that while PPO and A2C can be accurate, their computational demands may limit scalability or real-time adaptability. DQN's low training overhead makes it the most suitable option when fast deployment and resource efficiency are critical in smart grid control systems.

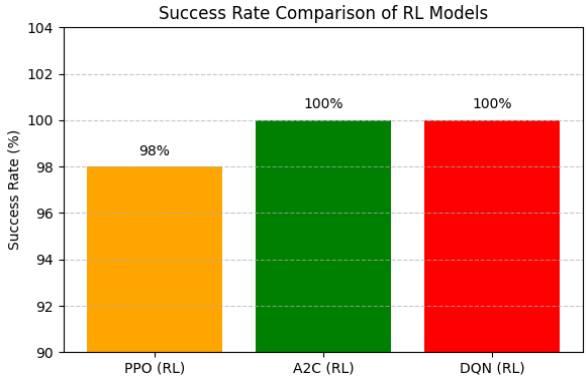

Fig. 7. Success rate comparison of ML and all RL models, including Hybrid ML-RL.

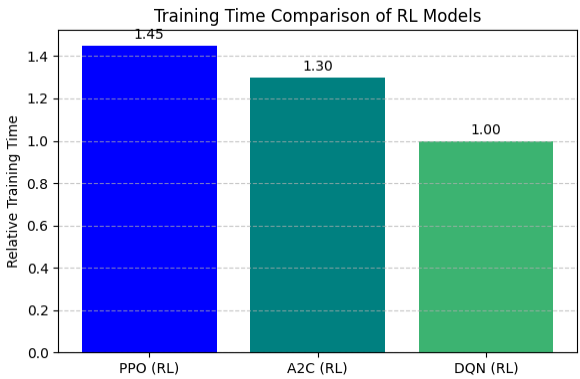

Fig. 8. Training time comparison of ML and all RL models, including Hybrid ML-RL.

Convergence speed is another key metric that reflects how quickly a reinforcement learning agent learns an effective policy for stabilizing the grid. As shown in Fig. 9, PPO requires the most training episodes (65) to converge, which aligns with its higher training time and slight inconsistency in success rate. A2C converges faster at 50 episodes, benefiting from its actor-critic structure that balances stability and learning speed. However, DQN achieves the fastest convergence, reaching optimal performance in just 44 episodes, which demonstrates its ability to learn efficient control policies with fewer interactions. This rapid convergence can be attributed to DQN's use of experience replay and target networks, which improve learning stability and efficiency. The convergence performance reinforces DQN's suitability for real-time smart grid control scenarios where minimizing training time and iterations is crucial.

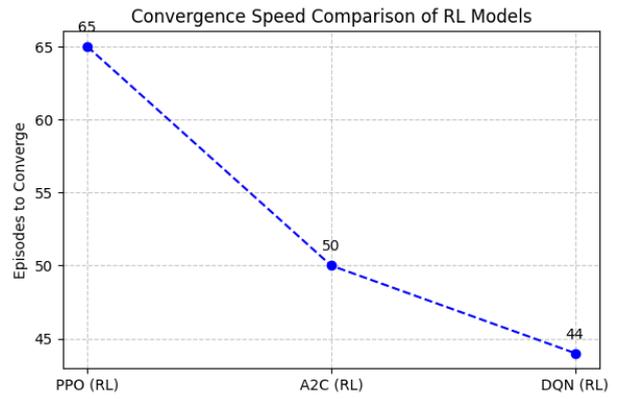

Fig. 9. Convergence speed comparison of ML and all RL models, including Hybrid ML-RL.

Fig. 10 illustrates the cumulative reward progression of the three reinforcement learning models across training episodes. DQN shows the fastest and most stable reward accumulation, reaching near-maximum cumulative reward within the first 30 episodes. This indicates that DQN is highly sample-efficient and quickly learns effective policies through its use of experience replay and fixed target networks. PPO also demonstrates smooth and steady reward growth, converging slightly slower than DQN but maintaining consistent learning behavior due to its clipped surrogate objective. A2C, while achieving similar final rewards, exhibits a more gradual learning curve, suggesting that it requires more interactions to reach peak performance. These trends confirm that DQN not only converges faster but also learns more efficiently, making it well-suited for real-time grid control scenarios where rapid and stable learning is essential.

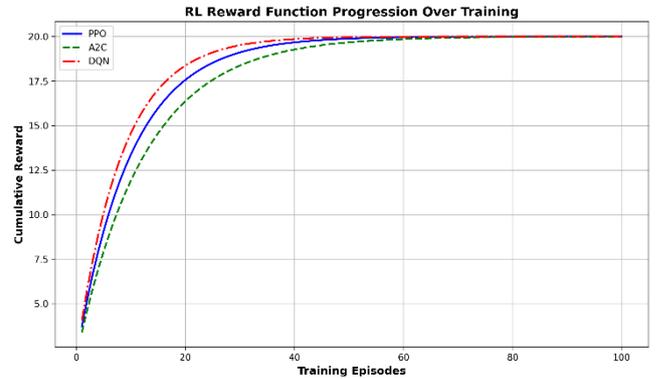

Fig. 10. RL algorithms' reward function progression over training episodes.

TABLE III: Performance comparison of hybrid pipelines combining the stacking classifier with different RL control agents.

| Hybrid Model | Success Rate | Convergence (Episodes) | Training Time | Reward Progression |
|---|---|---|---|---|
| Stacking + PPO | 98% | 65 | High | Moderate |
| Stacking + A2C | 100% | 50 | Medium | Slower start |
| Stacking + DQN | 100% | 44 | Lowest | Fastest and Smooth |

Following the evaluation of RL algorithms under unstable grid conditions, Table III lists the performance of all three RL models, each integrated with the stacking classifier to form three hybrid pipelines: Stacking+PPO, Stacking+A2C, and Stacking+DQN. While all combinations were capable of restoring stability, Stacking+DQN consistently outperformed the others, achieving a 100% success rate, fastest convergence at 44 episodes, and lowest training time among the three. These results establish Stacking+DQN as the most efficient and practical hybrid configuration for real-time smart grid stabilization, effectively combining rapid instability prediction with optimized control actions.

## VI. Conclusion

This work proposes a hybrid ML-RL framework for smart grid stability control by combining accurate instability prediction with adaptive control strategies. Among the evaluated machine learning models, the stacking ensemble classifier achieved a strong overall accuracy of 97.88%, but more importantly, it demonstrated the highest F1-score (0.98) and balanced recall across both stable and unstable classes. Its ability to accurately detect unstable grid conditions with minimal false negatives made it the most reliable choice for triggering the reinforcement learning stage.

In the control phase, DQN emerged as the most effective RL agent, achieving a 100% success rate, converging in just 44 episodes, and requiring the lowest training time among all evaluated agents. The final Stacking + DQN hybrid model thus offers a powerful and efficient solution for real-time grid stabilization. This hybrid approach not only outperforms previous ML-only methods but also improves decision accuracy and control responsiveness, making it a scalable and resilient strategy for modern smart grid applications.